\documentclass[showkeys,eqsecnum,twocolumn,showpacs,preprintnumbers,amssymb,aps,superscriptaddress]{revtex4}
\usepackage{graphicx}
\usepackage{dcolumn}
\usepackage{bm}
\usepackage{latexsym,epsfig}

\begin{document}

\preprint{\today}

\title{Comparative studies of dipole polarizabilities in Sr$^+$, Ba$^+$ and Ra$^+$ and their applications to optical clocks}
\vspace{0.5cm}

\author{B. K. Sahoo \footnote{B.K.Sahoo@rug.nl}, R. G. E. Timmermans}
\affiliation{KVI, University of Groningen, NL-9747 AA Groningen, The Netherlands}
\author{B. P. Das}
\affiliation{Non-accelerator Particle Physics Group, Indian Institute of Astrophysics, Bangalore-560034, India}
\author{D. Mukherjee}
\affiliation{Raman Center for Atomic, Molecular and Optical Sciences, IACS, Kolkata 70032, India}
\date{\today}
\vskip1.0cm

\begin{abstract}
Static dipole polarizabilities are calculated in the ground and metastable
states of
Sr$^+$, Ba$^+$ and Ra$^+$ using the relativistic coupled-cluster 
method. Trends of the electron correlation effects are investigated in these
atomic ions. We also estimate the Stark and black-body radiation
shifts from these results for these systems for the transitions proposed
for the optical frequency standards and compare them with available  
experimental data. 
\end{abstract}


\maketitle

\section{Introduction}
There have been a number of significant proposals for  
new optical clocks that are more accurate than the current standard; the Cs clock
\cite{diddams,adam,margolis}. Singly charged ions are some of the prominent 
candidates in this category due to the remarkable advances in modern ion trapping and laser cooling
techniques \cite{diddams,adam,margolis,itano1}. S-D transitions in 
Sr$^+$ \cite{margolis,madej}, Yb$^+$ \cite{schneider}, Hg$^+$ \cite{diddams,oskay}, Ba$^+$ \cite{sherman1,sherman2} and Ra$^+$ \cite{sahoo} can serve as clock transitions. It is necessary to estimate the shifts of the energy levels for 
these cases due to stray electromagnetic fields in order to determine the accuracies of these potential optical clocks. A knowledge of   
hyperfine structure constants, electric quadrupole moments, gyromagnetic
constants and polarizabilities are essential quantities that can be used
to estimate various possible shifts \cite{itano2}. In our previous studies, we have already
calculated hyperfine structure constants and electric quadrupole moments
for the above ions \cite{csur,bijaya1}. It is possible to find the gyromagnetic constants
approximately using analytical approaches for different states \cite{itano2}.
There have been extensive studies of the hyperfine structure 
constants of the low-lying states in the above ions using the relativistic 
coupled-cluster (RCC) method \cite{sahoo,bijaya1,bijaya2,bijaya3}. A few calculations of 
polarizabilities in these ions using the sum-over-states approach have
also been reported recently \cite{sahoo,jiang}. 

 Both Ba$^+$ and Ra$^+$ have also been proposed as suitable candidates
for atomic parity violation (APV) experiments \cite{fortson,wansbeek}. Determination of polarizabilities
depends on the electric dipole (E1) matrix elements and excitation energies.
On the otherhand, the determination of APV amplitudes also depends on E1 
matrix elements and excitation energies. Therefore, studies of correlation 
effects in these properties involving quantities in these systems are also useful for the APV studies. In contrast to
hyperfine structure constants where the explicit behavior of electron correlation
has been studied elaborately, the same cannot be done for polarizabilities 
using the sum-over-states approach. Also, the sum-over-states approach considers
only a limited number of states, mainly from the single excited states and
misses out contributions from continuum, double excited states,
normalization of the wave functions etc.

 In this work, we have employed an {\it ab initio} method in the RCC
framework to calculate dipole polarizabilities of Sr$^+$, Ba$^+$ and Ra$^+$.
The roles of different types of electron correlation effects in determining 
these quantities are studied and comparisons between these results are
given explicitly. Contributions arising from different
types of excited states and normalization of the wave functions 
through the RCC method have also been evaluated. Using these results, we then estimate 
the Stark and black-body radiation shifts in these systems which will be useful 
for the proposed optical clock experiments in the ions mentioned above.

\section{Theory}
The polarizability of a given state $| J_n M_n \rangle$ can be expressed by
\begin{eqnarray}
\alpha_0^i &=& 2 \sum_{m \ne n} C_i(J_n) \frac { |\langle J_n || D || J_m \rangle|^2}{E_n -E_m} \nonumber \\
&=& 2 \sum_{m \ne n} C_i(J_n) (-1)^{J_n-J_m} \frac { \langle J_n || D || J_m \rangle \langle J_m || D || J_n \rangle }{E_n -E_m},
\label{eqn1}
\end{eqnarray}
where the subscript $0$ represents for the static values and angular momentum 
coefficients ($C_i(J_n)$) for the scalar (with superscript 1) and tensor (with superscript 
2) dipole polarizabilities are given as 
\begin{eqnarray}
C_1(J_n) &=& - \frac{1}{3(2J_n+1)}, 
\label{eqn2}
\end{eqnarray}
and
\begin{eqnarray}
C_2(J_n) &=& \left [ \frac{10 J_n (2J_n-1)}{3 (J_n+1)(2J_n+1)(2J_n+3)} \right ]^2 \nonumber \\
&& \ \ \ \ \ \ \ \ (-1)^{J_n-J_m} \left \{ \matrix {J_n & 1 & J_m \cr 1 & J_n &
2 } \right \}, 
\label{eqn3}
\end{eqnarray}
respectively. $E$s are the energies of the corresponding atomic states.

By defining a modified wave function due to the dipole operator, $D$, we
can rewrite the above expression as
\begin{eqnarray}
 \alpha_0^i(J_n) &=& \langle J_n || \tilde{D}_i || J_n^{(1)} \rangle + \langle J_n^{(1)} || \tilde{D}_i || J_n \rangle,
\label{eqn4}
\end{eqnarray}
where 
\begin{eqnarray}
|J_n^{(1)} \rangle &=& \sum_{m \ne n} | J_m \rangle \frac {\langle J_m || D || J_n \rangle}{E_n -E_m}
\label{eqn5}
\end{eqnarray}
appears as a first order correction to the $|J_n \rangle$ state
due to the dipole operator $D$. In the above expression, we also define an
effective dipole operator as
\begin{eqnarray}
\tilde{D}_i=C_i(J_n) (-1)^{J_n-J_m} D ,
\label{eqn6}
\end{eqnarray}
whose matrix element between the original and perturbed wave functions will give the dipole polarizabilities.
Here the intermediate states, $| J_m \rangle$, have parities opposite to 
that of $| J_n \rangle$ and they have to satisfy the usual
triangular condition for the vector operator $D$.

To avoid the sum-over-states approach in the determination of the 
polarizabilities, we avoid the explicit form of $|J_n^{(1)} \rangle$
that is given by Eq. (\ref{eqn5}). In stead, we obtain $|J_n^{(1)} \rangle$ by solving the following equation
\begin{eqnarray}
(H-E_n) |J_n^{(1)} \rangle =  -D |J_n \rangle,
\label{eqn7}
\end{eqnarray}
that is similar to the first order perturbative equation. Here $H$ is the
atomic Hamiltonian which in the present work is considered in the Dirac-Coulomb approximation 
\begin{eqnarray}
H = \sum_i \left [ c \alpha \cdot p_i + (\beta -1)c^2 + V_{nuc}(r_i) \right ] + \sum_{i > j} \frac{1}{r_{ij}} ,
\label{eqn8}
\end{eqnarray}
where $c$ is the velocity of light, $\alpha$ and $\beta$ are the Dirac matrices
and $V_{nuc}(r)$ is the nuclear potential.

 For an atomic system with zero nuclear spin, the Stark shift to the second 
order in the presence of an electric field (quadratic Stark shift) for 
$|J_n, M_n \rangle$ state is given by \cite{angel}
\begin{eqnarray}
\Delta W_E(J_n, M_n; E)  &=& - \frac{1}{2} \alpha_0^1(J_n) E^2 - \frac{1}{4} \alpha_0^2(J_n) \nonumber \\ && \frac{[3M_n^2 -J_n (J_n+1)]}{J_n(2J_n-1)} (3E_z^2 -E^2), \nonumber \\ &&
\label{eqn9}
\end{eqnarray}
where $E$ and $E_z$ are the magnitudes of the externally applied electric 
field in any arbitrary and z directions, respectively.

 For atomic systems with non-zero nuclear spin ($I$), the expression for
hyperfine states are given by \cite{angel}
\begin{eqnarray}
\Delta W_E(F_n, M_{F_n}; E)  &=& - \frac{1}{2} \alpha_0^1(F_n) E^2 - \frac{1}{4} \alpha_0^2(F_n) \nonumber \\ && \frac{[3M_{F_n}^2 -F_n (F_n+1)]}{F_n(2F_n-1)} (3E_z^2 -E^2), \nonumber \\ &&
\label{eqn10}
\end{eqnarray}
where $F_n=I+J_n$ and $M_{F_n}$ are the total spin due to nuclear spin $I$ and 
atomic state spin $J_n$ and its azimuthal component, respectively. Since 
it is easier for us to deal with $J_n$ of the electronic states, therefore we 
express all the above quantities in terms of electronic coordinate. By using
the following relations \cite{itano2}
\begin{eqnarray}
\alpha_0^1(F_n) = \alpha_0^1(J_n)
\end{eqnarray}
and
\begin{eqnarray}
\alpha_0^2(F_n) &=& (-1)^{I+J_n+F_n} \alpha_0^2(J_n) \left [ \frac{F_n(2F_n-1)(2F_n+1)} {(2F_n+3)(F_n+1)} \right ] \nonumber \\ && \left [ \frac{(2J_n+1)(J_n+1)(2J_n+3)}{J_n(2J_n-1)} \right ] \left \{ \matrix { F_n & I & J_n \cr J_n & F_n & 2 \cr } \right \}, \nonumber \\ &&
\label{eqn11}
\end{eqnarray}
between the dipole polarizabilities of the electronic and hyperfine states,
we obtain
\begin{widetext}
\begin{eqnarray}
\Delta W_E(F_n, M_{F_n}; E)  &=& - \frac{1}{2} \alpha_0^1(J_n) E^2 - \frac{1}{4} (-1)^{I+J_n+F_n} \alpha_0^2(J_n) \left [ \frac{3M_{F_n}^2 -F_n (F_n+1)}{F_n(2F_n-1)} \right ] \nonumber \\ &&  \left [ \frac{F_n(2F_n-1)(2F_n+1)(2J_n+1)(J_n+1)(2J_n+3)}{(2F_n+3)(F_n+1)J_n(2J_n-1)} \right ] \left \{ \matrix { F_n & I & J_n \cr J_n & F_n & 2 \cr } \right \} (3E_z^2 -E^2).
\label{eqn12}
\end{eqnarray}
\end{widetext}

 Again, the blackbody-radiation (BBR) shift of a given state $|J_n, M_n \rangle$ in the adiabatic expansion 
due to the applied isotropic electric field radiated at temperature T (in 
Kelvin (K)) can be assumed as \cite{farley}
\begin{eqnarray}
\Delta_{BBR} = - \frac{1}{2} (831.9 V/m)^2 \left ( \frac{T(K)}{300} \right )^4 \alpha_0^1(J_n) .
\label{eqn13}
\end{eqnarray}

\section{Method of calculations}
The RCC method, which is equivalent to all order perturbation theory, has 
been recently used to obtain precise results and account for the correlation 
effects
accurately in single valence systems \cite{sahoo,csur,bijaya1,bijaya2,bijaya3}. In the RCC framework, the wave function of a single valence atom can be
expressed as 
\begin{eqnarray}
| \Psi_n^{(0)} \rangle &=& = e^T \{ 1+S_n \} | \Phi_n \rangle,
\label{eqn14}
\end{eqnarray}
where $| \Phi_n \rangle$ is the reference state constructed from the
Dirac-Fock wave function $| \Phi_0 \rangle$ of the closed-shell configuration
by appending the corresponding valence electron as $|\Phi_n \rangle= a_n^{\dagger} | \Phi_0 \rangle$ with $a_n^{\dagger}$ representing addition of a valence electron $n$. Here $T$ and $S_n$
are the RCC excitation operators which excite electrons from $| \Phi_0 \rangle$
and $|\Phi_n \rangle$, respectively. The amplitudes of these excitation 
are obtained by the following equations
\begin{eqnarray}
\langle \Phi^L |\{\widehat{H_Ne^T}\}|\Phi_0 \rangle = 0  \ \ \ \ \ \ \ \ \ \ \ \ \ \ \ \ \ \ \ \ \ \ \ \ \ \ \ \label{eqn14} \\
\langle \Phi_n^L|\{\widehat{H_Ne^T}\}S_n|\Phi_n\rangle = - \langle \Phi_n^L|\{\widehat{H_Ne^T}\}|\Phi_n\rangle \ \ \ \ \ \ \ \ \ \ \ \nonumber \\ \ \ \ \ \ \ \ \ \ \ \ \ \ \ \ + \langle \Phi_n^L|S_n|\Phi_n\rangle \Delta E_n , \ \ \ \ \ \
\label{eqn15}
\end{eqnarray}
with the superscript $L(=1,2)$ representing the single and double excited
states from the corresponding reference states and the wide-hat symbol over
$H_Ne^T$ represent the linked terms of normal order atomic Hamiltonian $H_N$
and RCC operator $T$. For the single and double excitations approximation
(CCSD method), the corresponding RCC operators are denoted by
\begin{eqnarray}
T &=& T_1 + T_2 
\label{eqn16}
\end{eqnarray}
and
\begin{eqnarray}
S_n &=& S_{1n} + S_{2n}
\label{eqn17}
\end{eqnarray}
for the closed-shell and single valence configurations, respectively. Again,
$\Delta E_n$ in the above expressions is the  corresponding valence electron
affinity (negative of the ionization potential (IP)) energy which is
evaluated by
\begin{eqnarray}
 \Delta E_n = \langle \Phi_n|\{\widehat{H_N e^T}\} \{1+S_n\} |\Phi_n\rangle.
\label{eqn18}
\end{eqnarray}
In Eq. (\ref{eqn14}), we have considered only
the single and double excitations, however we have incorporated contributions
from important triple excitations (CCSD(T) method) perturbatively in Eq. (\ref{eqn15}) by defining
\begin{eqnarray}
S_{3n}^{pert} &=&  \widehat{H_N T_2} + \widehat{H_N S_{2n}},
\label{eqn19}
\end{eqnarray}
where the superscript $pert$ denotes for the perturbation, and evaluating their contributions to $\Delta E_n$ from these operators by
\begin{eqnarray}
\Delta E_n^{trip} &=& \widehat{T_2^{\dagger} S_{3n}^{pert}} .
\label{eqn20}
\end{eqnarray}
After obtaining the amplitudes for $T$, the core excitation operator,
we solve Eqs. (\ref{eqn15}) and (\ref{eqn18}) simultaneously to obtain the
amplitudes for the $S_n$ operators.
\begin{table*}
\caption{Comparison of dipole polarizabilities between different works in Sr$^+$, Ba$^+$ and Ra$^+$.}
\begin{ruledtabular}
\begin{center}
\begin{tabular}{lccccccc}
System & ns$_{1/2}$ & \multicolumn{2}{c}{\underline{(n-1)d$_{3/2}$}} & \multicolumn{2}{c}{\underline{(n-1)d$_{5/2}$}} & Methods & References \\
 & $\alpha_0^1$  & $\alpha_0^1$ & $\alpha_0^2$ & $\alpha_0^1$ & $\alpha_0^2$ \\
\hline
 & & & & \\
 Sr$^+$($n=5$) & & & & \\
 & 127.62 & 145.86 & $-91.81$ & 136.84 & $-116.02$ & DF & This work \\
 & 88.29(1.0) & 61.43(52) & $-35.42(25)$ & 62.87(75) & $-48.83(30)$ & CCSD(T) & This work\\
 & 132.15  &  &  &         &            & HF & \cite{lim} \\
 &  86.21  &  &  &         &            & non-rel. MBPT(2) & \cite{lim} \\
 & 101.58  &  &  &         &            & non-rel. CCSD & \cite{lim} \\
 &  97.91  &  &  &         &            & non-rel. CCSD(T) & \cite{lim} \\
 & 121.33  &  &  &         &            & DK DF & \cite{lim} \\
 &  79.89  &  &  &         &            & DK rel. MBPT(2) & \cite{lim} \\
 &  94.31  &  &  &         &            & DK rel. CCSD & \cite{lim} \\
 &  91.10  &  &  &         &            & DK rel. CCSD(T) & \cite{lim} \\
 & 91.3(9) &  &  & 62.0(5) & $-47.7(3)$ & LCCSD(T)$+$sum-over & \cite{jiang} \\
 & 89.88 &  & & 61.77 &  & Non-rel.$+$sum-over & \cite{mitroy} \\
 & 93.3 &  & & 57.0 & &  Non-rel.$+$sum-over & \cite{barklem} \\
 & 84.6(3.6) &  & & 48(12) & & Non-rel.$+$sum-over & \cite{madej} \\
 & 91.47 &  & &  & & Non-rel.$+$sum-over & \cite{patil} \\
 & 86(11) & & &  & & Experiment + non-rel. & \cite{nunkaew} \\
 Ba$^+$($n=6$) & & & & \\
 & 184.49 & 90.07 & $-45.07$ & 87.66 & $-58.02$ & DF & This work \\
 & 124.26(1.0) & 48.81(46) & $-24.62(28)$ & 50.67(58) & $-30.85(31)$ & CCSD(T) & This work\\
 & 213.47  &  &  &         &            & HF & \cite{lim} \\
 & 110.60  &  &  &         &            & non-rel. MBPT(2) & \cite{lim} \\
 & 148.24  &  &  &         &            & non-rel. CCSD & \cite{lim} \\
 & 146.88  &  &  &         &            & non-rel. CCSD(T) & \cite{lim} \\
 & 177.64  &  &  &         &            & DK DF & \cite{lim} \\
 &  94.64  &  &  &         &            & DK rel. MBPT(2) & \cite{lim} \\
 & 129.92  &  &  &         &            & DK rel. CCSD & \cite{lim} \\
 & 123.07  &  &  &         &            & DK rel. CCSD(T) & \cite{lim} \\
 & 124.15  &  &  &         &         & LCCSD(T)$+$sum-over & \cite{iskrenova} \\
 & 124.7   &  &  &         &         & Non-rel.$+$sum-over & \cite{patil} \\
 & 126.2   &  &  &         &         & Non-rel.$+$sum-over & \cite{miadokova} \\
 & 123.88(5)  &  &  &         &      & Experiment & \cite{snow} \\
 & 125.5(10)  &  &  &         &      & Experiment & \cite{gallagher} \\
 Ra$^+$($n=7$) & & & & \\
 & 164.66 & 183.07 & $-114.70$ & 143.77 & $-98.64$ & DF & This work \\
 & 104.54(1.5) & 83.71(77) & $-50.23(43)$ & 82.38(70) & $-52.60(45)$ & CCSD(T) & This work\\
 & 257.00  &  &  &         &            & HF & \cite{lim} \\
 & 123.23  &  &  &         &            & non-rel. MBPT(2) & \cite{lim} \\
 & 186.23  &  &  &         &            & non-rel. CCSD & \cite{lim} \\
 & 172.00  &  &  &         &            & non-rel. CCSD(T) & \cite{lim} \\
 & 145.47  &  &  &         &            & DK DF & \cite{lim} \\
 &  79.80  &  &  &         &            & DK rel. MBPT(2) & \cite{lim} \\
 & 110.48  &  &  &         &            & DK rel. CCSD & \cite{lim} \\
 & 105.37  &  &  &         &            & DK rel. CCSD(T) & \cite{lim} \\
 & 106.12  &  &  &         &         & CCSD(T)$+$sum-over & \cite{sahoo} \\
 & 106.5  &  &  &         &         & LCCSD(T)$+$sum-over & \cite{safronova} \\
 & 106.22  &  &  &         &         & LCCSD(T)$+$sum-over & \cite{safronova1} \\
 & & & & \\
\end{tabular}
\end{center}
\end{ruledtabular}
\label{tab1}
Abbreviations: HF $\rightarrow$ Hartree-Fock. \\
 non-rel. $\rightarrow$ non-relativistic. \\ 
 DK rel. $\rightarrow$ scalar relativistic Douglas-Kroll method.\\ 
 MBPT(2) $\rightarrow$ second order perturbation theory.\\
 sum-over $\rightarrow$ sum-over intermediate states. 
\end{table*}

 We extend the RCC ansatz for the perturbed atomic state in the presence of 
the electric dipole operator $D$ by writing the total atomic wave function as
\begin{eqnarray}
| \tilde{\Psi}_n \rangle &=& = e^{T+ \Omega} \{ 1+S_n + \Lambda_n \} | \Phi_n \rangle,
\label{eqn21}
\end{eqnarray}
where $\Omega$ and $\Lambda_n$ are the first order corrections to the RCC operators 
$T$ and $S_n$, respectively. Since Eq. (\ref{eqn7}) is first order in
the operator $D$, the above expression will reduce to
\begin{eqnarray}
| \tilde{\Psi}_n \rangle &=& = e^T \{ 1+ S_n+ \Omega ( 1+S_n) + \Lambda_n \} | \Phi_n \rangle .
\label{eqn22}
\end{eqnarray}
Now, separating the above wave function as $| \Psi_n^{(0)} \rangle$ and $| \Psi_n^{(1)} \rangle$, we get
\begin{eqnarray}
| \Psi_n^{(1)} \rangle &=& = e^T \{ \Omega ( 1+S_n) + \Lambda_n \} | \Phi_n \rangle .
\label{eqn23}
\end{eqnarray}
Following Eq. (\ref{eqn7}), we solve again the amplitudes for the modified 
operators as
\begin{eqnarray}
\langle \Phi^L |\{\widehat{H_Ne^T} \Omega \}|\Phi_0 \rangle &=&  - \langle \Phi^
L | \widehat{De^T} |\Phi_0 \rangle   \label{eqn24} \\
\langle \Phi_n^L|\{\widehat{H_Ne^T}\}\Lambda_n|\Phi_n\rangle &=& - \langle 
\Phi_n^L|\{\widehat{H_Ne^T} \Omega ( 1+S_n) + \widehat{De^T} \nonumber \\ && ( 1+S_n)
 \}|\Phi_n\rangle + \langle \Phi_n^L|\Lambda_v|\Phi_n\rangle \Delta E_n , \nonumber \\
\label{eqn25}
\end{eqnarray}
where $\widehat{De^T}$ represents the connecting terms between $D$ and $T$
operators. Again in our CCSD approximation, we have
\begin{eqnarray}
\Omega &=& \Omega_1 + \Omega_2 
\label{eqn26}
\end{eqnarray}
and
\begin{eqnarray}
\Lambda_n &=& \Lambda_{1n} + \Lambda_{2n} .
\label{eqn27}
\end{eqnarray}

Therefore, the RCC expression for the dipole polarizability is given by
\begin{widetext}
\begin{eqnarray}
\alpha_0^i &=& \frac {\langle \Psi_n^{(0)} | \tilde{D}_i | \Psi_n^{(1)} \rangle
+ \langle \Psi_n^{(1)} | \tilde{D}_i | \Psi_n^{(0) } \rangle} {<\Psi_n^{(0) }|\Psi_n^{(0) }>} \nonumber \\
&=& \frac {\langle \Phi_n |\{1+S_n^{\dagger}\} \overline{\tilde{D}_i} \{\Omega(1
+S_n) + \Lambda_n \} | \Phi_n \rangle + \langle \Phi_n |\{\Lambda_n^{\dagger} +
(1+S_n^{\dagger}) \Omega^{\dagger} \} \overline{\tilde{D}_i} \{1 +S_n\} | \Phi_n\rangle } { \langle \Phi_n | \{1+S_n^{\dagger}\} \overline{N}_0 \{1 +S_n\} | \Phi_n \rangle },
\end{eqnarray}
\label{eqn28}
\end{widetext}
where we define $\overline{\tilde{D}_i}=(e^{T^{\dagger}} \tilde{D}_i e^T)$ and
$\overline{N}_0 = e^{T^{\dagger}} e^T$. The non-truncative series for
$\overline{\tilde{D}_i}$ and $\overline{N}_0$ are expanded using the Wick's 
generalized theorem and truncated the series when the leading order 
non-accounted terms are below fifth order of
Coulomb interaction. These operators are then contracted with the $\Omega$
to get fully contracted terms that give rise core electron contributions.
The core-valence and valence correlation contributions are obtained from
the open contraction between the operators with $\Omega$ and $\Omega\{1+S_n\}+\Lambda_n$ operators, respectively.

Corrections due to the normalization of the wave functions are accounted by 
evaluating
\begin{eqnarray}
Norm &=& \left [ \langle \Psi_n^{(0)} | \tilde{D}_i | \Psi_n^{(1)} \rangle + \langle \Psi_n^{(1)} | \tilde{D}_i | \Psi_n^{(0)} \rangle \right ]\{ \frac {1}{1+N_n} - 1 \}, \nonumber \\ &&
\label{eqn29}
\end{eqnarray}
where $N_n=\langle \Phi_n | \{1+S_{n}^{\dagger}\} \overline{N}_0 \{1 +S_{n}\}  | \Phi_n \rangle $.

\section{Results and Discussions}

\subsection{General discussions}
\begin{figure}[h]
\includegraphics[width=8.5cm,clip=true]{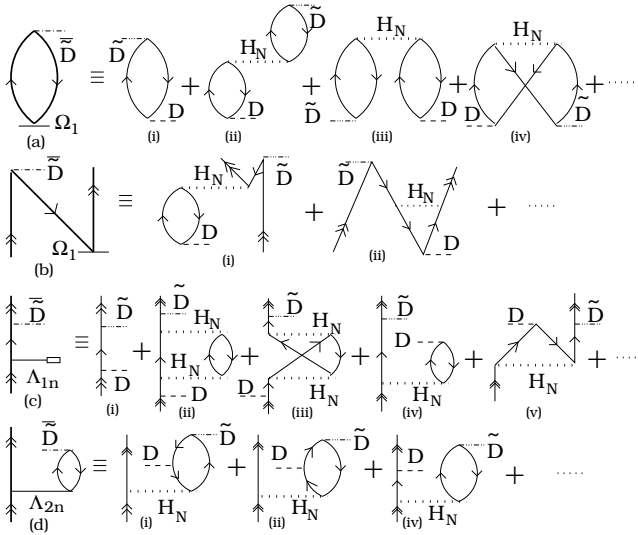}
\caption{Break-down of RCC terms into lower-order perturbative diagrams.}
\label{fig1}
\end{figure}

The orbitals used in the present work are generated on a radial grid given by
\begin{eqnarray}
r_i = r_0 \left [ e^{h(i-1)} - 1 \right ],
\label{eqn28}
\end{eqnarray}
where $i$ represents a grid point. The total number of grid points in our calculations is $750$, 
the step size $h$ is taken as $0.03$ in the present case and $r_0$ is
taken as $ 2 \times 10^{-6}$ atomic units. To construct the basis
functions, we use Gaussian type orbitals (GTOs) defined as
\begin{eqnarray}
F^{GTO}(r_i) = r^{n_{\kappa}} e^{-\alpha_i r_i^2} .
\label{eqn29}
\end{eqnarray}
Here $n_{\kappa}$ is the radial quantum number of the orbitals
and $\alpha_i$ is a parameter whose value is chosen to
obtain orbitals with proper behavior inside and outside the nucleus of
an atomic system. Further, the $\alpha_i$s satisfy the even tempering condition
\begin{eqnarray}
\alpha_i = \alpha_0 \beta^{i-1}.
\label{eqn30}
\end{eqnarray}

\begin{table*}
\caption{Contributions from the DF and various CCSD(T) terms to the dipole polarizability calculations in Sr$^+$. The subscripts $c$, $n$ and $v$ of the RCC terms correspond to the core, valence and virtual correlation contributions. $\tilde{D}_c$ and $\tilde{D}_n$ give the DF results from the core and valence orbitals.}
\begin{ruledtabular}
\begin{center}
\begin{tabular}{lccccc}
Terms  & 5s $^2S_{1/2}$  & \multicolumn{2}{c}{4d $^2D_{3/2}$}  & \multicolumn{2}{c}{4d $^2D_{5/2}$} \\
 & $\alpha_0^1$ & $\alpha_0^1$ & $\alpha_0^2$ & $\alpha_0^1$ & $\alpha_0^2$ \\
\hline \\
$\tilde{D}_c$ & 6.15 & 6.15 & $-0.25$ & 6.15 & $-0.25$ \\
$\tilde{D}_n$ & 121.47 & 139.71 & $-91.56$ & 130.69 & $-115.78$ \\
 & & \\
$\overline{\tilde{D}_c}$ & 4.98 & 4.98 & $-0.27$ & 4.98 & $-0.27$ \\
$\overline{\tilde{D}_v} \Omega + cc$ & 0.10 & 0.20 & $-0.35$ & 0.41 & $-0.41$ \\
$\overline{\tilde{D}_v} \Lambda_{1n} + cc$ & 93.78 & 67.05 & $-38.70$ & 68.14 & $-56.54$ \\
$\overline{\tilde{D}_v} \Lambda_{2n} + cc$ & $-2.87$ & $-2.54$ & $-0.86$ & $-2.43$ & 1.72 \\
$S_{1v} \overline{\tilde{D}_v} \Lambda_{1n} + cc$ & $-3.74$ & $-6.06$ & 3.47 & $-6.10$ & 5.05 \\
$S_{2v} \overline{\tilde{D}_v} \Lambda_{1n} + cc$ & $-3.03$  & $-1.47$ & 0.77 & $-1.39$ & 0.96 \\
$S_{1v} \overline{\tilde{D}_v} \Lambda_{2n} + cc$ & $-0.04$  & $0.06$ & 0.03 & 0.05 & $-0.03$ \\
$S_{2v} \overline{\tilde{D}_v} \Lambda_{2n} + cc$ & $0.13$ & $-0.05$ & 0.03 & $-0.06$ & 0.03 \\
Others & 0.13 & 0.22 & $-0.14$ & 0.24 & $-0.15$ \\
$Norm$ & $-1.15$ & $-0.96$ & 0.60 & $-0.97$ & 0.81 \\
\end{tabular}
\end{center}
\end{ruledtabular}
\label{tab2}
\end{table*}
We have chosen same $\alpha_0=0.00525$ and $\beta=2.73$ values to construct
the basis functions in Sr$^+$, Ba$^+$ and Ra$^+$, so that effects due to the
different sizes of the systems can be compared. Finite size of the nucleus in these
systems are accounted by assuming a two-parameter Fermi-nuclear-charge 
distribution for evaluating the electron density over nucleus as given by
\begin{equation}
\rho(r_i) = \frac {\rho_0} {1 + e^{(r_i-c)/a}}, \label{eqn26}
\label{eqn31}
\end{equation}
where $c$ and $a$ are the half-charge radius and skin thickness of the 
nucleus. These parameters are chosen as 
\begin{equation}
a = 2.3/4(ln3)
\end{equation}
and
\begin{equation}
c = \sqrt{ \frac{5}{3} r_{rms}^2 - \frac{7}{3} a^2 \pi^2},
\label{eqn32}
\end{equation}
where $r_{rms}$ is the root mean square radius of the corresponding nuclei 
which is determined as discussed in \cite{johnson}.

 In Table \ref{tab1}, we present our DF and CCSD(T) results along with 
other available calculations and experimental results for the
dipole polarizabilities of the ground and metastable states
of Sr$^+$, Ba$^+$ and Ra$^+$. 
The differences between the DF and CCSD(T) results indicate the magnitudes 
of the electron correlation effects in the determination of the dipole 
polarizabilities in these systems using the CCSD(T) method.
They are 45\%, 48\% and 58\% for the ground states of
Sr$^+$, Ba$^+$ and Ra$^+$, respectively. They increase with
the size of the system. However, the correlation effects
in the $d_{3/2}$ metastable states are 137\%, 85\% and 119\% for 
the scalar polarizabilities and 159\%, 83\% and 128\% for the tensor
polarizabilities in Sr$^+$, Ba$^+$ and Ra$^+$, respectively. 
This shows that the correlation effects reduce in these states from 
Sr$^+$ to Ba$^+$, but the presence of the core f-orbitals increases
the correlation effects in Ra$^+$. The correlation effects
in the $d_{5/2}$ metastable states are 117\%, 73\% and 75\% for the
scalar polarizabilities and 138\%, 88\%  and 87\% for the tensor
polarizabilities in Sr$^+$, Ba$^+$ and Ra$^+$, respectively. This
implies that the correlations in the d-metastable states 
do not depend upon the size but the internal structure of the systems.
Our previous studies on the hyperfine structure
constants in these systems \cite{sahoo,bijaya1,bijaya2,bijaya3} had 
shown peculiar behavior of the core-polarization effects. 
These effects were comparatively smaller in the
$d_{3/2}$ metastable states. In contrast, the correlation 
effects are larger in the $d_{3/2}$ metastable states compared
to the $d_{5/2}$ metastable states and the ground states in the dipole
polarizabilities calculations.
\begin{table*}
\caption{Contributions from the DF and various CCSD(T) terms to the dipole polarizability calculations in Ba$^+$. The subscripts $c$, $n$ and $v$ of the RCC terms correspond to the core, valence and virtual correlation contributions. $\tilde{D}_c$ and $\tilde{D}_n$ give the DF results from the core and valence orbitals.}
\begin{ruledtabular}
\begin{center}
\begin{tabular}{lccccc}
Terms  & 6s $^2S_{1/2}$  & \multicolumn{2}{c}{5d $^2D_{3/2}$}  & \multicolumn{2}{c}{5d $^2D_{5/2}$} \\
 & $\alpha_0^1$ & $\alpha_0^1$ & $\alpha_0^2$ & $\alpha_0^1$ & $\alpha_0^2$ \\
\hline \\
$\tilde{D}_c$ & 11.73 & 11.73 & $-0.46$ & 11.73 & $-0.46$ \\
$\tilde{D}_n$ & 172.76 & 78.33 & $-44.61$ & 75.93 & $-57.56$ \\
 & & \\
$\overline{\tilde{D}_c}$ & 9.35 & 9.35 & $-0.56$ & 9.35 & $-0.56$ \\
$\overline{\tilde{D}_v} \Omega + cc$ & 0.23 & 0.33 & $-0.64$ & 0.82 & $-0.82$ \\
$\overline{\tilde{D}_v} \Lambda_{1n} + cc$ & 133.01 & 49.20 & $-25.61$ & 50.17 & $-36.23$ \\
$\overline{\tilde{D}_v} \Lambda_{2n} + cc$ & $-4.93$ & $-3.23$ & $-1.36$ & $-3.05$ & 2.19 \\
$S_{1v} \overline{\tilde{D}_v} \Lambda_{1n} + cc$ & $-6.58$ & $-4.45$ & 2.39 & $-4.42$ & 3.37 \\
$S_{2v} \overline{\tilde{D}_v} \Lambda_{1n} + cc$ & $-5.18$  & $-1.98$ & 0.84 & $-1.83$ & 0.89 \\
$S_{1v} \overline{\tilde{D}_v} \Lambda_{2n} + cc$ & $-0.06$  & $0.05$ & 0.03 & 0.03 & $-0.01$ \\
$S_{2v} \overline{\tilde{D}_v} \Lambda_{2n} + cc$ & $0.27$ & $-0.10$ & 0.05 & $-0.09$ & 0.06 \\
Others & 0.21 & 0.45 & $-0.25$ & 0.49 & $-0.33$ \\
$Norm$ & $-2.06$ & $-0.81$ & 0.49 & $-0.80$ & 0.59 \\
\end{tabular}
\end{center}
\end{ruledtabular}
\label{tab3}
\end{table*}

 The upper limits to the error bars in these quantities were determined by 
taking the differences of the results obtained using CCSD(T) and CCSD
methods and the inaccuracies due to the self-consistent 
results obtained at the DF levels by varying the number of GTOs considered in the
calculations. These results are quoted inside the 
parentheses in Table \ref{tab1}.

 We explicitly present the diagrams in Fig. \ref{fig1} corresponding to
various RCC terms that are significant in determining the 
dipole polarizabilities. As seen from the figure,
Fig. \ref{fig1}(a) which arises from the fully contracted terms of $\overline{\tilde{D}}\Omega_1$ corresponds to the core-correlation contributions. Its lower
order terms corresponds mainly to the diagrams coming from the random-phase 
approximation (RPA). There are also core-correlation contributions arising from $\overline{\tilde{D}}\Omega_2$, but they are relatively small and are not 
shown in Fig. \ref{fig1}. The core-valence correlation contributions are 
determined by open diagrams from $\overline{\tilde{D}}\Omega_1$ as shown
in Fig. \ref{fig1}(b). The most important correlation contributions arise 
through the valence correlation effects and they are shown in Fig. \ref{fig1}(c)
 and \ref{fig1}(d). Important pair-correlation and core-polarization effects
are accounted through $\overline{\tilde{D}}\Lambda_{1n}$, however
core-polarization effects arising from the perturbed doubly excited states
are accounted through $\overline{\tilde{D}}\Lambda_{2n}$. The DF 
contributions involving the core, core-valence and virtual orbitals
are the lowest order diagrams to the fully contracted $\overline{\tilde{D}}\Omega_1$, open $\overline{\tilde{D}}\Omega_1$ and $\overline{\tilde{D}}\Lambda_{1n}$
RCC terms, respectively. Based on the above mentioned correlation diagrams, we  
analyze their roles in different systems considered below.

\subsection{Sr$^+$}
There are no experimental results of the dipole polarizabilities available 
for the ground and metastable excited states in Sr$^+$. However, a number of 
calculations have been carried out using different  
methods and we have compared their results with the present work in 
Table \ref{tab1}. Lim and Schwerdtfeger \cite{lim} have done  comparative 
studies between the non-relativistic and scalar relativistic Douglas-Kroll 
calculations using four different many-body methods. They demonstrate the 
importance of the relativistic methods to calculate dipole polarizabilities.
Jiang et al. \cite{jiang} have used E1 matrix elements
obtained using the linearized RCC method with the singles, doubles and partial
triple excitations (LCCSD(T)) to evaluate the valence correlation 
contributions for a few intermediate states. The core-correlations are accounted 
through the RPA method and contributions from higher states were estimated 
using the DF method. Mitroy et al. \cite{mitroy} have used a non-relativistic
method using the sum-over-states approach to determine polarizabilities of the
ground and d-state. As seen in Table \ref{tab1}, the dipole polarizabilities
of the 4d$_{3/2}$ and 4d$_{5/2}$ states are not the same and they cannot be
evaluated separately using a non-relativistic method. However, our 
ground state polarizability for Sr$^+$ agrees with their result. Similar
approaches were also employed by Barklem and O’Mara \cite{barklem}.
Patil and Tang \cite{patil} have employed a summation and integration approach 
to determine the ground state polarizability. Recently, Nunkaew et al.
\cite{nunkaew} have estimated E1 matrix elements using the non-relativistic theory
and microwave resonance measurements in Sr and have extracted dipole polarizability of the ground state of Sr$^+$.
\begin{table*}
\caption{Contributions from the DF and various CCSD(T) terms to the dipole polarizability calculations in Ra$^+$. The subscripts $c$, $n$ and $v$ of the RCC terms correspond to the core, valence and virtual correlation contributions. $\tilde{D}_c$ and $\tilde{D}_n$ give the DF results from the core and valence orbitals.}
\begin{ruledtabular}
\begin{center}
\begin{tabular}{lccccc}
Terms  & 7s $^2S_{1/2}$  & \multicolumn{2}{c}{6d $^2D_{3/2}$}  & \multicolumn{2}{c}{6d $^2D_{5/2}$} \\
 & $\alpha_0^1$ & $\alpha_0^1$ & $\alpha_0^2$ & $\alpha_0^1$ & $\alpha_0^2$ \\
\hline \\
$\tilde{D}_c$ & 15.56 & 15.56 & $-0.56$ & 15.56 & $-0.56$ \\
$\tilde{D}_n$ & 149.10 & 167.51 & $-114.14$ & 128.21 & $-98.07$ \\
 & & \\
$\overline{\tilde{D}_c}$ & 11.66 & 11.66 & $-0.71$ & 11.66 & $-0.71$ \\
$\overline{\tilde{D}_v} \Omega + cc$ & 0.60 & 0.21 & $-0.54$ & 1.03 & $-1.03$ \\
$\overline{\tilde{D}_v} \Lambda_{1n} + cc$ & 107.74 & 91.30 & $-54.32$ & 85.59 & $-62.17$ \\
$\overline{\tilde{D}_v} \Lambda_{2n} + cc$ & $-4.15$ & $-5.92$ & $-2.84$ & $-4.85$ & 3.68 \\
$S_{1v} \overline{\tilde{D}_v} \Lambda_{1n} + cc$ & $-5.17$ & $-7.62$ & 4.51 & $-6.65$ & 5.13 \\
$S_{2v} \overline{\tilde{D}_v} \Lambda_{1n} + cc$ & $-4.93$  & $-4.64$ & 2.68 & $-3.44$ & 1.76 \\
$S_{1v} \overline{\tilde{D}_v} \Lambda_{2n} + cc$ & $-0.02$  & $0.05$ & 0.05 & 0.02 & $-0.01$ \\
$S_{2v} \overline{\tilde{D}_v} \Lambda_{2n} + cc$ & $0.25$ & $-0.18$ & 0.07 & $-0.15$ & 0.10 \\
Others & 0.29 & 0.61 & $-0.34$ & 0.64 & $-0.43$ \\
$Norm$ & $-1.73$ & $-1.76$ & 1.21 & $-1.47$ & 1.08 \\
\end{tabular}
\end{center}
\end{ruledtabular}
\label{tab4}
\end{table*}

In Table \ref{tab2}, we present the individual contributions from RCC terms 
to the dipole polarizability calculations in Sr$^+$. Our core-correlation contributions are 4.98 au and $-0.27$ au for the scalar and tensor 
dipole polarizabilities, respectively. Clearly, the CCSD(T) result for the 
scalar dipole polarizability is smaller than the previously estimated values. 
On the otherhand, the core-correlation to the tensor polarizability vanishes 
in the non-relativistic theory, but it is finite in our approach, although 
small in magnitude. Jiang et al. \cite{jiang} have neglected this contribution
in their calculations. We have also given DF results from the core 
($\tilde{D}_c$) and virtual ($\tilde{D}_v$) orbitals separately in the same
table. Our DF result and that reported by Lim and Schwerdtfeger \cite{lim} differ.
Comparing our DF results given in Table \ref{tab1} and Table \ref{tab2}, it 
seems that Lim and Schwerdtfeger have not included core correlation 
contributions at the DF level. Again, the lowest order contributions to 
$\overline{\tilde{D}_c}$ and $\overline{\tilde{D}_v} \Lambda_{1n}$ terms 
correspond to $\tilde{D}_c$ and $\tilde{D}_v$, respectively. The
differences between the lowest order and
all order results seem to be significant in this system. The largest contributions to the final results come from $\overline{\tilde{D}_v} \Lambda_{1n}$ as it 
contains DF results due to virtual orbitals in it. Contributions from $\overline{\tilde{D}_v} \Lambda_{2n}$ correspond to doubly excited perturbed states and they are also large
in both the ground and metastable states. Therefore, the exclusion of these
contributions in the sum-over-state approach may not be appropriate. Again,
normalization corrections ($Norm$) are also non-negligible.

\subsection{Ba$^+$}
Two experimental results with small uncertainties \cite{snow,gallagher} are available for the ground state dipole polarizability in Ba$^+$. There 
have also been studies of this quantity by Lim and Schwerdtfeger
\cite{lim}. Iskrenova-Tchoukova and Safronova \cite{iskrenova} have employed
E1 matrix elements from the LCCSD(T) method in the sum-over-states approach 
using a few states for the valence correlation effects  and 
estimating the core-correlation and core-valence correlation contributions 
from lower order perturbation theory to determine this quantity. Other 
available calculations \cite{patil,miadokova} are based on 
non-relativistic methods. Again, there are no other results available
for the metastable d-states in Ba$^+$ to compare with our results. However,
we have also carried-out a sum-over-states calculation using the E1 matrix
elements from the CCSD(T) method \cite{bijaya4} that agrees with our 
{\it ab initio} results.

We present contributions from individual RCC terms to the 
dipole polarizabilities calculations on Ba$^+$ in Table \ref{tab3}. The trends of these 
correlation effects seem to be the same as in Sr$^+$. However, the core correlation effects in this system seem to be almost twice than in the case of Sr$^+$. The core-valence 
correlations coming through the open $\overline{\tilde{D}_v} \Omega$
diagrams are also larger than Sr$^+$. Contributions
from the doubly excited perturbed states and corrections due to the 
normalization of the wave functions also seem to be significant.

\subsection{Ra$^+$}
There are also no experimental results available for the dipole 
polarizabilities in Ra$^+$. In the same work as mentioned above, Lim and 
Schwerdtfeger \cite{lim} have also calculated this quantity in the ground 
state of Ra$^+$ using various many-body methods. Safronova et al 
\cite{safronova,safronova1} have also evaluated this result using the sum-over-states
approach. Their valence correlation effects are evaluated using E1 matrix 
elements for a few important states from the LCCSD(T) method  and 
core-correlation and core-valence correlations are evaluated using 
lower order many-body methods. In our earlier work \cite{sahoo}, we had also
evaluated dipole polarizabilities in the ground and d-metastable states
using the sum-over-states approach with the E1 matrix elements from CCSD(T)
method and approximated core-correlation and core-valence correlation 
effects. 

In Table \ref{tab4}, we present contributions from individual RCC terms
to these results. The trend of the correlation effects in the ground state
seems similar to those of Sr$^+$ and Ba$^+$, but due to the presence of 
core f-electrons, the behavior of the correlation effects is a 
little different for the metastable d-states. The size of core-correlation 
is slightly larger than that of Ba$^+$, but the difference is not as large as it was
between Sr$^+$ and Ba$^+$. 
In contrast to Ba$^+$ where the {\it ab initio} and
sum-over-states results match, we found discrepancies in this system.
The discrepancies are mainly because of the inclusion of the doubly excited states
in the present work, but there could be cancellations in Ba$^+$ due to which 
the discrepancies are small.

\subsection{Applications to the optical clocks}
All the ions considered in this work are important candidates for optical 
clocks \cite{margolis,madej,sherman1,sherman2,sahoo}. 
There has been an absolute frequency measurement of the 5s $^2S_{1/2} \rightarrow $ 4d $^2D_{5/2}$ transition in $^{88}$Sr$^+$ by Madej et al. \cite{madej}. 
One of the largest uncertainties
due to the applied electric field comes from the quadratic Stark shift. 
In fact, this shift was earlier over estimated
due to the large error bars in the calculated dipole polarizabilities 
of the 5s $^2S_{1/2}$ and 4d $^2D_{5/2}$ states. Madej et al. 
had used $\alpha_0^1(5s_{1/2})=(1.40\pm0.06)\times 10^{-39}$ C$^2$ s$^2$ 
kg$^{-1}$ where we obtain this result as $(1.46\pm0.02)\times 10^{-39}$ C$^2$ 
s$^2$ kg$^{-1}$. The scalar and tensor polarizabilities of the 4d $^2D_{5/2}$
were used in \cite{madej} as $\alpha_0^1(4d_{5/2})=(8\pm2)\times 10^{-40}$ 
C$^2$ s$^2$ kg$^{-1}$ and $\alpha_0^2(4d_{5/2})=(-7\pm2)\times 10^{-40}$ 
C$^2$ s$^2$ kg$^{-1}$, respectively. We obtain these results as 
$\alpha_0^1(4d_{5/2})=(10.37\pm0.12)\times 10^{-40}$ C$^2$ s$^2$ kg$^{-1}$ and
$\alpha_0^2(4d_{5/2})=(-8.05\pm0.05)\times 10^{-40}$ C$^2$ s$^2$ kg$^{-1}$.
Using Eq. (\ref{eqn9}) and our results, we obtain the shift rate, which is
defined as $\gamma= \frac {\delta (\Delta W_E)}{\delta E^2}$, of the 
5s $^2S_{1/2}$ state as $(1.10\pm0.01)$ $\mu$Hz/(V/m)$^2$ against $(1.06\pm0.04) \ \mu$Hz/(V/m)$^2$ of
Madej et al. Similarly, we obtain $\gamma=(-0.78\pm0.02) \ \mu$Hz/(V/m)$^2$ 
against results of Madej et al. as $\gamma=(-0.6\pm0.2) \ \mu$Hz/(V/m)$^2$  
in the 4d $^2D_{5/2}$ state using only the scalar polarizability. However
assuming the direction of the electric field lies in the z-direction, 
we obtain $\gamma=(-1.27\pm0.03) \ \mu$Hz/(V/m)$^2$, $\gamma=(-0.91\pm0.02) \ \mu$Hz/(V/m)$^2$ and $\gamma=(-0.18\pm0.01) \ \mu$Hz/(V/m)$^2$ for $M=1/2$, $M=3/2$ and $M=5/2$, respectively.

 Using Eq. (\ref{eqn13}) and the above results, we also obtain the 
black-body radiation shift at $T=300$K in the 5s $^2S_{1/2} \rightarrow $ 
4d $^2D_{5/2}$ transition in $^{88}$Sr$^+$ as $(0.22\pm0.01)$ Hz and that is 
an improvement of 10\% over the result of Madej et al. \cite{madej}. 

 It appears that both $^{137}$Ba$^+$ and $^{138}$Ba$^+$ will be suitable 
candidates for an optical clock \cite{sherman1,sherman2}, but each has some 
advantages and disadvantages in controlling the systematic errors. For 
the 6s $^2S_{1/2} \rightarrow $ 5d $^2D_{5/2}$ transition in 
$^{138}$Ba$^+$, it would be possible to use techniques similar to the measurement of the frequency in the optical transition in $^{88}$Sr$^+$ mentioned 
earlier. However, one has to encounter the electric quadrupole 
shift in the 5d $^2D_{5/2}$ state for this case. It is possible to 
overcome this particular shift by considering the possible $F=2(6s_{1/2}) \rightarrow F=0(5d_{3/2})$ hyperfine transition in $^{137}$Ba$^+$. In this
transition, one has to again estimate the possible quadratic Zeeman shifts
because of finite nuclear magnetic and quadrupole moments. Our dipole
polarizability for the 6s $^2S_{1/2}$ state is
given by $(2.05\pm0.02)\times 10^{-39}$ C$^2$ s$^2$ kg$^{-1}$.
The scalar and tensor polarizabilities of the 5d $^2D_{3/2}$ state
are given by $(8.05\pm0.07)\times 10^{-40}$ C$^2$ s$^2$ kg$^{-1}$ and
$(-4.06\pm0.05)\times 10^{-40}$ C$^2$ s$^2$ kg$^{-1}$, respectively.
Similarly, the scalar and tensor polarizabilities of the 5d $^2D_{5/2}$ state
are given by $(8.35\pm0.09)\times 10^{-40}$ C$^2$ s$^2$ kg$^{-1}$ and
$(-5.09\pm0.05)\times 10^{-40}$ C$^2$ s$^2$ kg$^{-1}$, respectively. Due 
to the choice of the hyperfine transition in $^{137}$Ba$^+$,
the tensor polarizabilities of these states are zero
and hence the polarizabilities of the atomic and hyperfine states are the same.
The shift rates are $(-1.55\pm0.01) \ \mu$Hz/(V/m)$^2$ and
$(-0.61\pm0.01) \ \mu$Hz/(V/m)$^2$ in the 6s $^2S_{1/2}$ and  5d $^2D_{3/2}$
states, respectively. For $^{138}$Ba$^+$, by considering particular $M$
values of the 5d $^2D_{5/2}$ state and assuming that the electric field lies in the z-direction, we can evaluate the Stark shifts. They 
are $\gamma=(-0.94\pm0.02) \ \mu$Hz/(V/m)$^2$, $\gamma=(-0.71\pm0.02)
\ \mu$Hz/(V/m)$^2$ and $\gamma=(0.14\pm0.01) \ \mu$Hz/(V/m)$^2$ for $M=1/2$, $M
=3/2$ and $M=5/2$, respectively. As can be noticed, the result for $M=5/2$ has opposite sign than other $M$ values. The Stark shifts in these states can be easily estimated
using these results for a given applied electric field.

The black-body radiation shift at $T=300$K in the 6s $^2S_{1/2} \rightarrow $
5d $^2D_{5/2}$ transition in this system is given as $(0.64\pm0.12)$ Hz. 
 
 Similarly as we had reported earlier \cite{sahoo}, both $^{223}$Ra$^+$ and $^{225}$Ra$^+$
have the same advantages like $^{137}$Ba$^+$ and $^{138}$Ba$^+$, respectively, 
for considering
as optical clock candidates. In fact, all the low-lying energy 
levels in these ions are in optical region which will be an advantage for
the experimentalists to measure the 7s $^2S_{1/2} \rightarrow $ 6d $^2D_{3/2}$  or 7s $^2S_{1/2} \rightarrow $ 6d $^2D_{5/2}$ or $F=2(7s_{1/2}) \rightarrow F=0(6d_{3/2})$ transition frequencies more precisely than other candidates. Recently, 
$^{213}$Ra whose half-lifetime is around 2.75 $m$ was produced at KVI \cite{shidling}
in the accelerator method and its single ion shares the same advantage with 
$^{225}$Ra$^+$ for becoming suitable candidate for the optical clock. Now assuming
that due to the suitable choice of hyperfine states in $^{223}$Ra$^+$ \cite{sahoo} like the case for $^{137}$Ba$^+$, the tensor polarizability contribution to the Stark-shift will be zero and hence using
our dipole polarizability results, we obtain the Stark shift rates as 
$(-1.31\pm0.02) \ \mu$Hz/(V/m)$^2$ and $(-1.05\pm0.02) \ \mu$Hz/(V/m)$^2$ in the 7s $^2S_{1/2}$ and 6d $^2D_{3/2}$ states, respectively. 
For other isotopes discussed above, by considering particular $M$
values of the 6d $^2D_{5/2}$ state and assuming that the electric field lies in the z-direction, the Stark shifts are evaluated as
$\gamma=(-1.56\pm0.03) \ \mu$Hz/(V/m)$^2$, $\gamma=(-1.16\pm0.02)
\ \mu$Hz/(V/m)$^2$ and $\gamma=(0.29\pm0.01) \ \mu$Hz/(V/m)$^2$ for $M=1/2$, $M
=3/2$ and $M=5/2$, respectively. The result for $M=5/2$ has opposite sign than other $M$ values like in $^{138}$Ba$^+$. Therefore, the Stark shifts in Ra$^+$ can be estimated accurately using our results for a given applied electric field.

The black-body radiation shift at $T=300$K in the 7s $^2S_{1/2} \rightarrow $
6d $^2D_{5/2}$ transition in Ra$^+$ is given as $(0.19\pm0.02)$ Hz.

 From the above Stark shift ratios and BBR shifts in the considered ions, 
it is found that these systematic errors are small in Ra$^+$ which further
supports along with its energy level locations that it will be one of 
the most suitable candidates for optical clock. In fact, a possible atomic 
clock with uncertainty in the order of $10^{-17}$ seems feasible 
from these results along with the preliminary analysis of Doppler's shifts
\cite{jungmann} in Ra$^+$.

\section{Conclusion}
We have employed the relativistic coupled-cluster method to determine
{\it ab initio} results for the dipole polarizabilities of the ground
and the metastable d-states in the singly
ionized strontium, barium and radium. Electron correlation effects arising
through various coupled-cluster terms are given individually and  
comparative studies are performed for these three ions. Using the results
we have obtained, Stark shifts and black-body radiation shifts for these
ions are estimated. Using our results, we were able to reduce the errors
of the measured frequency for the optical clock in $^{88}$Sr$^+$. Our 
calculations of the Stark and black-body radiations shifts in both 
Ba$^+$ and Ra$^+$ could be used to remove the systematic errors in the
proposed optical clock experiments for these ions.

\section{Acknowledgment}
This work was supported by NWO under the VENI program
with Project No. 680-47-128 and part of the Stichting
FOM Physics Program 48 TRI$\mu$p.
We thank the C-DAC TeraFlop Super Computing facility, Bangalore, India for the cooperation to carry
out these calculations on its computers.

\end{document}